\documentclass[a4paper,12pt]{article}

\usepackage{epsf}
\usepackage{amsmath}
\usepackage{amssymb}
\usepackage{graphicx}
\usepackage{graphics}
\usepackage{dcolumn}
\usepackage{bm}
\usepackage{color}
\usepackage{nohyperref}
\usepackage{cite}
\usepackage{soul}

\begin{document}
\begin{center}
\Large\bf\boldmath
\vspace*{0.8cm} Measuring $V_{ub}$ and
probing SUSY with double ratios of purely leptonic decays
of $B$ and $D$ mesons
\\
\unboldmath
\end{center}
\vspace{0.6cm}
\begin{center}
A.G. Akeroyd$^{a,}$\footnote{Electronic address: \tt akeroyd@ncu.edu.tw} and F. Mahmoudi$^{b,}$\footnote{Electronic address: 
\tt mahmoudi@in2p3.fr} \\[0.4cm]
\vspace{0.6cm}
{\sl a: Department of Physics, National Central University, Jhongli, Taiwan 320}\\
{\sl b: Clermont Universit\'e, Universit\'e Blaise Pascal, CNRS/IN2P3,\\ 
LPC, BP 10448, 63000 Clermont-Ferrand, France}
\end{center}

\vspace{0.7cm}
\begin{abstract}
\noindent The experimental prospects for precise measurements of the leptonic decays $B_u\to \tau\nu/\mu\nu$, $B_s\to \mu^+\mu^-$, $D\to \mu\nu$ and $D_s\to \mu\nu/\tau\nu$ are very promising. Double ratios involving four of these decays can be defined in which the dependence on the values of the decay constants is essentially eliminated, thus enabling complementary measurements of the CKM matrix element $V_{ub}$ with a small theoretical error.
We quantify the experimental error in a possible future measurement of $|V_{ub}|$ using this approach, and show that it is competitive with the anticipated precision from the conventional approaches. Moreover, it is shown that such double ratios can be more effective than the individual leptonic decays as a probe of the parameter space of supersymmetric models. We emphasize that the double ratios have the advantage of using $|V_{ub}|$ as an input parameter (for which there is experimental information), while the individual decays have an uncertainty from the decay constants (e.g. $f_{B_s}$), and hence a reliance on theoretical techniques such as lattice QCD.
\end{abstract}

\vspace{0.3cm}

\section{Introduction}

Measurements of the Cabibbo-Kobayashi-Maskawa (CKM) matrix elements 
\cite{Cabibbo:1963yz,Kobayashi:1973fv} are of great importance,
being fundamental parameters of the Standard Model (SM). The matrix elements display a strong hierarchy, and the smaller off-diagonal terms can be expressed as higher powers of an expansion parameter 
$\lambda(\sim 0.22)$ \cite{Wolfenstein:1983yz}. 
One of the smallest elements is $V_{ub}$, being of order $\lambda^3$ and with a magnitude $\sim 4\times 10^{-3}$.
Although the measurement of $|V_{ub}|$ is very challenging, in recent years its precision  
has been reduced to $\sim 10\%$ at the $B$ factories. 
Two ways to measure $|V_{ub}|$ have been utilized: inclusive semi-leptonic decays
($B\to X_u \ell \nu$) and exclusive semi-leptonic decays ($B\to \pi\ell \nu$). These two approaches
provide independent measurements of $|V_{ub}|$, which are currently not in total agreement, but have roughly the same precision.
Both of these methods can be employed at high-luminosity $B$ factories, and
the experimental error in the determination of $|V_{ub}|$ will decrease with increasing
integrated luminosity and improved analysis techniques. The inclusive and exclusive determinations
of $|V_{ub}|$ have independent theoretical errors, and it is acknowledged that a concurrent reduction of this theoretical error (i.e. comparable in degree to that for the experimental error)
is more challenging (e.g. see \cite{Amsler:2008zz}). Therefore, as emphasized in 
\cite{Grinstein:1993ys,Ligeti:1995yz,Ligeti:2003hp,Grinstein:2009qv}
it is important to consider alternative methods of measuring $|V_{ub}|$ for which the theoretical error is known to be very small, even if the experimental error for such a method is currently large.

One such approach is the branching ratio (BR) of the purely leptonic decay $B_u\to \tau\nu$.
The unprecedented data samples provided by the $B$ factories have enabled the first measurements of 
 $B_u\to \tau\nu$ \cite{Ikado:2006un, Adachi:2008ch, Aubert:2007bx,Aubert:2007xj}
despite its relatively small BR and challenging signature. The current world average of 
BR($B_u\to \tau\nu$) has an error of $20\%$, although proposed high-luminosity flavour factories could reach a precision of $5\%$ 
\cite{Yamauchi:2002ru,Bigi:2004kn,Hashimoto:2004sm,Akeroyd:2004mj,Browder:2004wu,Hewett:2004tv,Bona:2007qt,Browder:2007gg,Browder:2008em}.
A measurement of $|V_{ub}|f_B$ can be obtained from this decay, where $f_B$ is the decay constant for the $B_u$ meson, which has been calculated by Lattice Quantum Chromodynamics (LQCD). 
An average of two unquenched calculations \cite{Bernard:2009wr,Gamiz:2009ku} of $f_B$ gives $f_B=192.8\pm 9.9$ MeV \cite{Laiho:2009eu}, i.e. a precision of $\sim 5\%$. Hence the decay $B_u\to \tau\nu$ (and its companion decay $B_u\to \mu\nu$, which is yet to be observed), could provide complementary (and possibly competitive) measurements of $|V_{ub}|$ at high-luminosity flavour factories.

However, despite the impressive improvement in the precision of calculations of $f_B$ by LQCD, it is important to extract $|V_{ub}|$ with minimal theoretical error (i.e. with minimal dependence on parameters like $f_B$, which are determined by the strong interaction).
It was pointed out long ago \cite{Grinstein:1993ys} that a double ratio of the leptonic decays 
$B_u\to \tau\nu, B_s\to \mu^+\mu^-, D\to \mu\nu$ and $D_s\to \mu\nu/\tau\nu$ 
has essentially no dependence on the absolute values of the decay constants.
The decay constants in this double ratio appear as $(f_{B_s}/f_B)/(f_{D_s}/f_D)$, and such a quantity is numerically very close to 1, with a very small error. Hence such a double ratio can be used to obtain a
measurement of $|V_{ub}|$  with a small theoretical error \cite{Ligeti:2003hp}. 
At the time of writing of \cite{Grinstein:1993ys}, only one of the four decays had been measured 
($D_s\to \mu\nu/\tau\nu$). At present, three of the four decays have been observed, and
the prospect of precise measurements of all four decays in this double ratio 
motivates us to quantify the expected precision for a measurement of $|V_{ub}|$ using this method.

The decays $D\to \mu\nu$ \cite{:2008sq}, $D_s\to \mu\nu$ 
\cite{Alexander:2009ux,:2007ws,Aubert:2006sd} and $D_s\to \tau\nu$ 
\cite{Alexander:2009ux,Onyisi:2009th,Collaboration:2009tk,:2010qj}
have been measured with $\sim 10\%$ precision at CLEO-c and the $B$ factories, and the ongoing 
BES-III experiment will reduce the error in all these decays to maybe $3\%$ \cite{Li:2008wv,Asner:2008nq}.
The current measurements of $B_u\to \tau\nu$, $D\to \mu\nu$, $D_s\to \mu\nu$ and $D_s\to \tau\nu$ are in agreement with the SM prediction, within the errors of the decay constants. An upper limit exists for the decay $B_s\to\mu^+\mu^-$ ($< 4.3 \times 10^{-8}$ at 95 C.L. \cite{CDF9892}) and this is the only decay in the double ratio which has not been observed. At the CERN LHC, the LHCb experiment is expected to achieve sensitivity to the SM prediction of BR($B_s\to\mu^+\mu^-$) $\sim 3\times 10^{-9}$. Therefore the first measurement of the double ratio will be possible by combining the measurements of these four leptonic decays from three distinct experimental facilities.

The above purely leptonic decays are sensitive to New Physics particles which arise in supersymmetric (SUSY) models, such as charged Higgs bosons ($H^\pm$), neutral Higgs bosons $(h^0, H^0$ and $A^0)$, and charginos ($\chi^\pm$).
The decays $B_u\to \tau\nu$, $D\to \mu\nu$, $D_s\to \mu\nu$ and $D_s\to \tau\nu$
are sensitive to $H^\pm$ at the {\sl tree level} \cite{Hou:1992sy}. 
The largest effect is for $B_u\to \tau\nu$ and the current measurement of this BR, although in agreement with the SM prediction, does not preclude a large contribution from $H^\pm$\cite{Eriksson:2008cx}\footnote{However, any sizeable contribution of $H^\pm$ on $B_u\to \tau\nu$ is now possibly disfavoured by considering the contribution of $H^\pm$ on other decays such as $B \to D\tau\nu$.}.
 The effect of  $H^\pm$ on 
$D_s\to \mu\nu$ and $D_s\to \tau\nu$ is much smaller \cite{Hou:1992sy, Hewett:1995aw}, but can be of the order of the current experimental precision \cite{Akeroyd:2003jb,Akeroyd:2007eh,Akeroyd:2009tn}.
The contribution of $H^\pm$ to $D\to \mu\nu$ is negligible. 
The decay $B_s\to\mu^+\mu^-$, which proceeds via higher-order diagrams in perturbation theory, can be greatly enhanced by contributions from neutral Higgs bosons and chargino-stop loops. The contributions from SUSY grow with $\tan\beta$, which can lead to an enhancement of BR($B_s\to\mu^+\mu^-$) by an order of magnitude with respect to the SM expectation \cite{Choudhury:1998ze,Babu:1999hn}.

Since the decays involved in the leptonic double ratios can receive substantial contributions from Higgs bosons and supersymmetric particles, another possible application of the double ratio is to constrain SUSY parameters. The absence of an explicit dependence on the decay constants makes the double ratio a particularly attractive quantity with which to probe SUSY. 
In this case, $V_{ub}$ is an input parameter and is the only major source of uncertainty. 
Information on $V_{ub}$ is available from both direct measurements (the aforementioned inclusive and exclusive semileptonic decays, the latter having some dependence on LQCD) and from global SM fits. In contrast, the calculation of decay constants is reliant on LQCD. 
Consequently, due to its reduced theoretical uncertainties the double ratio constitutes a cleaner observable than the individual decays, and can provide competitive constraints on SUSY parameters. 

The paper is organized as follows: in section 2 the conventional ways to measure $|V_{ub}|$ are briefly summarized and the double ratio of leptonic decays is introduced. 
In section 3, the decay rates of the leptonic decays are presented and the experimental prospects for their measurements are discussed.
The current upper limit on the double ratio is derived, as well as the anticipated error in a measurement of $|V_{ub}|$ at future experimental facilities using the double ratio.  
Section 4 contains a numerical study of the double ratio as a probe of the SUSY parameter space
in two distinct models. Conclusions are given in section 5.

%%%%%%%%%%%%%%%%%%%%%%%%%%%%%%%%%%%%%%%%%%%%%%%%%%%%%%%%%%%%%%%%%%%%
%
\section{Measurements of $|V_{ub}|$}
Currently there are two methods which are used to measure $|V_{ub}|$.
Both methods use a semi-leptonic decay channel of the $B/B_u$ meson for 
which the decay rate depends on $|V_{ub}|^2$. 
One method (``inclusive approach'') is to measure the branching ratio of the inclusive decay 
$B\to X_u\ell \nu$ while the other method (``exclusive approach'')
is to measure the branching ratio of the exclusive decays $B\to \pi\ell \nu$ and  $B\to \rho\ell \nu$.
The value of $|V_{ub}|$ is then extracted from these measurements, with an additional error from the theoretical calculation of the decay rate, which is independent for the two decays. Currently there is a small disagreement between the values of $|V_{ub}|$ extracted from these distinct approaches.

%%%%%%%%%%%%%%%%%%%%%%%%%%%%%%%%%%%%%%%%%%%%%%%%%%%%%%%%%%%%%%%%%%%%
%
\subsection{Inclusive determination of $|V_{ub}|$}
The main difficulty with the inclusive determination of $|V_{ub}|$
is the fact that the decay $B\to X_c\ell \nu$ has a branching ratio
about fifty times larger than that of $B\to X_u\ell \nu$ (because $|V_{cb}|\gg|V_{ub}|$). In order to reduce this background and isolate the signal from  $B\to X_u\ell \nu$, kinematical cuts are
applied. One approach is to select charged leptons $\ell$ with greatest energy (the ``endpoint region'').
For an energy $E_{\ell}> 2.4$ GeV the background from $B\to X_c\ell \nu$
is very small, but the theoretical prediction for the decay rate of $B\to X_u\ell \nu$ in this endpoint region has sizeable errors, which are summarized in \cite{Amsler:2008zz}. The most significant error is 
from the non-perturbative ``shape function''\cite{Neubert:1993ch}. Fortunately, the leading shape function is universal in $B/B_u$ decays. Therefore the dependence of the decay rate of $B\to X_u\ell \nu$ (in the endpoint region) on the leading shape function can be eliminated by using data from other $B$ decays (such as $B\to X_s\gamma$). 
 However, ``sub-leading shape functions'' break this universality of the
shape function, and so the process of eliminating its effect on $B\to X_u\ell \nu$ brings in additional errors. 
In \cite{Grinstein:2009qv} it is suggested that this error from sub-leading shape functions might limit the ultimate precision in such measurements of $|V_{ub}|$ to around $15\%$.
Alternatively, the dependence of the decay rate of $B\to X_u\ell \nu$ on the shape function can be reduced by applying a less energetic cut on $\ell$ (e.g.  $E_{\ell} > 1.9$ GeV), but the background from $B\to X_c\ell \nu$ then becomes much larger.

There are several measurements of $|V_{ub}|$ from the inclusive method, and different 
kinematical cuts are applied. In order to reduce the background from
 $B\to X_c\ell \nu$, most measurements select only a small fraction (partial rate) of the
total inclusive rate of $B\to X_u\ell \nu$. The value of $|V_{ub}|$ is then extracted
in the context of several theoretical prescriptions, which treat the various theoretical errors 
(including the dependence on the shape function) in distinct ways.

A recent measurement by the BELLE Collaboration \cite{:2009tp}
applies a much lower cut on the energy of $\ell \;($$E_{\ell} > 1.0$ GeV), which keeps $\sim 90\%$ of the total inclusive rate of $B\to X_u\ell \nu$.
Such an approach substantially reduces the theoretical error (especially that from the shape function)
in the extraction of $|V_{ub}|$, and it is suggested in 
\cite{Akeroyd:2004mj} that this method is the most promising one 
for a precise measurement of $|V_{ub}|$ from inclusive  $B\to X_u\ell \nu$ at a high-luminosity $B$ factory.

The current measurement of  $|V_{ub}|$ from inclusive decays uses 
eight independent measurements of $B\to X_u\ell \nu$, and is an average over the results of the four 
theoretical prescriptions \cite{Kowalewski}:
\begin{equation}
|V_{ub}|=4.37\pm 0.39\times 10^{-3} \;.
\label{vub-inclusive}
\end{equation}
In Eq.(\ref{vub-inclusive}) the theoretical error and experimental error have been combined, 
and the experimental error ($\pm 0.16$) is smaller than the theoretical error.

%%%%%%%%%%%%%%%%%%%%%%%%%%%%%%%%%%%%%%%%%%%%%%%%%%%%%%%%%%%%%%%%%%%%
%
\subsection{Exclusive determination of $|V_{ub}|$}

The decays $B\to \pi\ell \nu$ and  $B\to \rho\ell \nu$ have both been measured, but the former is much more favourable for extracting $|V_{ub}|$, both experimentally and theoretically.
There are two search strategies for  $B\to \pi\ell \nu$. One technique tags the other $B/B_u$ in the event, while the other technique does not. 
Currently these techniques have comparable precisions, but it is suggested that the tagged method
(e.g. as used by BELLE in \cite{Hokuue:2006nr}) will afford the greatest precision at a high-luminosity $B$ factory.

Extracting the value of $|V_{ub}|$ requires theoretical input distinct from that used in
the inclusive approach.  The decay $B\to \pi\ell \nu$ only depends on one hadronic form factor
$f_+(q^2)$ (where $q$ is the momentum of $\ell$) which can be calculated in LQCD for $q^2> 16$ GeV$^2$ 
and by Soft Collinear Effective Theory (SCET) for $q^2< 16$ GeV$^2$.
The world average measurement of $|V_{ub}|$ from exclusive decays is \cite{Laiho:2009eu}:
\begin{equation}
|V_{ub}|=3.42\pm 0.37\times 10^{-3} \;.
\label{vub-exclusive}
\end{equation}
A recent BABAR measurement alone \cite{:2010uj} gives $|V_{ub}|=2.95\pm 0.31\times 10^{-3}$.
Therefore both the inclusive and exclusive measurements of $|V_{ub}|$ have comparable
experimental and theoretical errors, but there is a disagreement of the central values.
However, the average of the inclusive and exclusive measurements is $|V_{ub}|=(3.92\pm0.09\pm0.45)\times10^{-3}$ \cite{Charles:2004jd},
which agrees quite well with the values of $|V_{ub}|$ extracted from two independent global SM fits, the latter being $|V_{ub}|=3.53\pm 0.15\times 10^{-3}$ \cite{Charles:2004jd} and $|V_{ub}|=3.52\pm 0.11\times 10^{-3}$ \cite{Bona:2009cj}.

%%%%%%%%%%%%%%%%%%%%%%%%%%%%%%%%%%%%%%%%%%%%%%%%%%%%%%%%%%%%%%%%%%%%
%
\subsection{The double ratio $(f_B/f_{B_s})/(f_D/f_{D_s})$}
Although the experimental prospects for both the inclusive and exclusive approaches to measure
$|V_{ub}|$ are good, the question of the magnitude of the theoretical error will remain. 
A complementary technique for which the theoretical error is known to be small is particularly attractive.
In this regard, the use of double ratios of leptonic decays to measure $|V_{ub}|$ was proposed in \cite{Grinstein:1993ys,Ligeti:2003hp}.
One such double ratio is defined by:
\begin{equation}
\frac{\Gamma(B_u\to\tau\nu)}{\Gamma(B_s\to \mu^+\mu^-)}
\frac{\Gamma(D_s\to\mu\nu)}{\Gamma(D\to \mu\nu)}\sim
\frac{|V_{ub}|^2}{|V_{ts}V_{tb}|^2}\;\frac{\pi^2}{\alpha^2}\;
\frac{f_B/f_{B_s}}{f_D/f_{D_s}} \;.
\label{doub-rat}
\end{equation}

The quantity $(f_B/f_{B_s})/(f_D/f_{D_s})$ deviates from unity by small corrections
of the form $m_s/m_b$ and $m_s/m_c$, and importantly such a double ratio 
does not have a dependence on the absolute values of the
decay constants.
A calculation in \cite{Grinstein:1993ys} gives $(f_B/f_{B_s})/(f_D/f_{D_s})=0.967$, and subsequent works 
\cite{Oakes:1994tj} also give values very close to 1, with a very small error. In our numerical analysis we will take $(f_B/f_{B_s})/(f_D/f_{D_s})=1$.
Therefore Eq.~(\ref{doub-rat}) can be used to measure $|V_{ub}|$ with essentially no dependence
on non-perturbative techniques such as LQCD. This amounts to using 
the decay $B_u\to \tau\nu$ to measure $|V_{ub}|$, with the dependence on $f_B$ (i.e. the dependence on non-perturbative techniques) being removed by forming the double ratio of decay constants. In Eq.~(\ref{doub-rat}), the product of CKM elements $|V_{ts}V_{tb}|^2$ is
known with high precision if one assumes that the CKM matrix is unitary.
The left-hand side of Eq.~(\ref{doub-rat}) is taken from experiment, and thus
a measurement of $|V_{ub}|$ can be extracted when all four decays have been measured. 
The drawback of this approach is that accurate measurements of all four decays
are required.  At the time of writing of \cite{Grinstein:1993ys} only one decay
had been measured. At present, three of the decays in Eq.~(\ref{doub-rat}) have been measured, 
while an upper limit exists for $B_s\to \mu^+\mu^-$.
Evidently, the double ratio cannot yet provide a measurement of $|V_{ub}|$
because one of the decays has not been measured. However, the experimental prospects
for reasonably precise measurements of all four decays are quite good, and
eventually a measurement of $|V_{ub}|$ using Eq.~(\ref{doub-rat}) will be possible.

Other double ratios can be defined, such as replacing $D_s\to\mu\nu$ in Eq.~(\ref{doub-rat})
by $D_s\to \tau\nu$, both of these decays having comparable experimental precision at present 
(and similar precision is expected for both decays at the BES-III experiment \cite{Asner:2008nq}).
Moreover, the decay $B_u\to \mu\nu$ is expected to be observed at a high-luminosity $B$ factory, which would provide an independent measurement of $|V_{ub}|f_B$, and enable a measurement of a double ratio in which $B_u\to \tau\nu$ is replaced by  $B_u\to \mu\nu$.
We note that a SM-like rate for the decay $D\to\tau\nu$ will be measured at the BES-III experiment, but its precision will always be inferior to that for $D\to\mu\nu$ \cite{Zweber:2009qf}, and so it is preferable to use this latter decay in the definition of double ratios.

In the context of models beyond the SM, the double ratio is sensitive to New Physics particles. Assuming that the partial widths for each decay are multiplied by scale factors, the right-hand side of the double ratio in Eq.~(\ref{doub-rat}) is modified by a ratio of scale factors:
\begin{equation}
\frac{r_B \;r_{D_s}}{r_{B_s}}\equiv R^{-1} \;.
\end{equation}
In the next section we will give the explicit form of each of the scale factors appearing in $R$, in the context of the Minimal Supersymmetric Standard Model (MSSM).

%%%%%%%%%%%%%%%%%%%%%%%%%%%%%%%%%%%%%%%%%%%%%%%%%%%%%%%%%%%%%%%%%%%%
%
\section{The leptonic decays in the double ratio}
In following subsections we present the theoretical expressions for the decay rates of the 
leptonic decays in the double ratio, and discuss the experimental prospects for their measurements.
The current upper bound on the double ratio is derived by combining statistically the existing measurements of the individual decays.
Finally, the attainable experimental precision for $|V_{ub}|$ from a measurement of 
the double ratio at the next generation of experiments is estimated.

%%%%%%%%%%%%%%%%%%%%%%%%%%%%%%%%%%%%%%%%%%%%%%%%%%%%%%%%%%%%%%%%%%%%
%
\subsection{The decay $B_u\to \tau\nu$}
In the SM this decay is mediated by $W^\pm$ and is helicity suppressed, whereas there is no such suppression for the scalar $H^\pm$ exchange, whose contribution is proportional to the $b$ quark and $\tau$ lepton Yukawa couplings. 
In the limit of high $\tan\beta$ such Yukawa couplings are enhanced, and
the contribution from $H^\pm$ can be comparable in magnitude to that of $W^\pm$ 
\cite{Hou:1992sy}. The leading order SM prediction for this decay is:
\begin{equation}
\mathrm{BR}(B_u\to\tau\nu_\tau)_\mathrm{SM}=\frac{G_F^2f_B^2|V_{ub}|^2}{8\pi\Gamma_B}m_Bm_\tau^2\left(1-\frac{m_\tau^2}{m_B^2}\right)^2\;,
\label{eq:Btaunu}
\end{equation}
while the New Physics contribution from $H^\pm$ is expressed through the ratio 
\cite{Hou:1992sy,Akeroyd:2003zr,Itoh:2004ye}
\begin{equation}
\label{eq:RMSSM}
r_B \equiv \frac{\mathrm{BR}(B_u\to\tau\nu_\tau)_{\mathrm{MSSM}} }{\mathrm{BR}(B_u\to\tau\nu_\tau)_{\mathrm{SM}}}=\left[1-\left(\frac{m_B^2}{m_{H^+}^2}\right)\frac{\tan^2\beta}{1+\epsilon_0\tan\beta}\right]^2\;.
\end{equation}
Here $m_{H^+}$ is the
mass of the charged Higgs boson, $m_{B}$ is the mass of the $B_u$ meson, $\Gamma_B$ is the total decay width of the $B_u$ meson, $\tan\beta=v_2/v_1$ where $v_1$ and $v_2$ are the vacuum expectation values of the two scalar doublets, and $G_F$ is the Fermi constant.   
The leading SUSY-QCD corrections are included in this expression through $\epsilon_0$.
The Yukawa couplings of the MSSM take the form of a 2HDM (Type~II) at tree level, but at higher orders the structure becomes of the type 2HDM (Type III) in which $\epsilon_0$ is a function of SUSY parameters
\cite{Hall:1993gn,Hempfling:1993kv,Carena:1994bv,Blazek:1995nv} and $|\epsilon_0|$ can reach values of order 0.01.
Using the average $f_B=192.8\pm9.9$ MeV \cite{Laiho:2009eu} of two unquenched lattice QCD calculations
of $f_B$ \cite{Bernard:2009wr,Gamiz:2009ku}, and the average of the inclusive and exclusive determinations of $|V_{ub}|=(3.92\pm0.09\pm0.45)\times10^{-3}$ \cite{Charles:2004jd}, we evaluate the SM branching ratio with SuperIso v2.8 \cite{Mahmoudi:2007vz,Mahmoudi:2008tp}:
\begin{equation}
\mathrm{BR}(B_u\to\tau\nu_\tau)_\mathrm{SM}= (1.01\pm 0.26)\times 10^{-4}\;.
\end{equation}
There are four independent measurements of $B_u\to \tau\nu$ \cite{Ikado:2006un,Adachi:2008ch,Aubert:2007bx,Aubert:2007xj}
and the current world average is  \cite{Barberio:2008fa}:
\begin{equation}
\mathrm{BR}(B_u \to \tau\nu_\tau)_\mathrm{exp}=  (1.63\pm 0.39)\times 10^{-4}\;. \label{interval_btaunu}
\end{equation}
The SM prediction can be compared to the experimental average by forming the ratio:
\begin{equation}
r_B^{\mathrm{exp}} \equiv \frac{\mathrm{BR}(B_u\to\tau\nu_\tau)_{\mathrm{exp}} }{\mathrm{BR}(B_u\to\tau\nu_\tau)_{\mathrm{SM}}} =1.62\pm 0.57\;.
\label{Rtaunu} 
\end{equation}

Significant improvements in the precision of the measurement of $B_u\to \tau\nu$
will require a high-luminosity $B$ factory. The two measurements by BABAR \cite{Aubert:2007bx,Aubert:2007xj} have used a large amount ($\sim 70\%$) of the available data taken at $\Upsilon (4S)$.  
The two measurements by BELLE \cite{Ikado:2006un,Adachi:2008ch} are with 414 fb$^{-1}$
and 605 fb$^{-1}$, and so these could be significantly updated with the total integrated luminosity 
of 1000 fb$^{-1}$.
A high-luminosity $B$ factory with around 50 ab$^{-1}$ could measure $B_u\to \tau\nu$ to a precision of
around $6\%$ (e.g. see  \cite{Bona:2007qt}). Moreover, with 50 ab$^{-1}$ the decay $B_u\to \mu\nu$ 
(for which there is currently an upper limit) could be measured with about the same precision as $B_u\to \tau\nu$. Hence a precision of around $3\%$ for $|V_{ub}|f_B$ could achieved from each decay at a high-luminosity $B$ factory. A combination of the measurements of $|V_{ub}|f_B$ from $B_u\to \tau\nu$ and $B_u\to \mu\nu$ would (presumably) further reduce the uncertainty. 

%%%%%%%%%%%%%%%%%%%%%%%%%%%%%%%%%%%%%%%%%%%%%%%%%%%%%%%%%%%%%%%%%%%%
%
\subsection{The decays $D_s \to \tau\nu$, $D_s \to \mu\nu$ and $D \to \mu\nu$}
In analogy to the case for $B_u\to \tau\nu$, singly charged Higgs bosons
would also contribute to the decays  $D_s \to \ell\nu$ and $D \to \ell\nu$ at tree level \cite{Hou:1992sy}.
The effect is negligible for  $D \to \ell\nu$, but it can be of the order of the current experimental
precision for  $D_s \to \mu\nu$ and $D_s \to \tau\nu$.
The partial width is given by (where $\ell=e,\mu$ or $\tau$):
\begin{equation}
\Gamma(D_s\to \ell\nu_\ell) = \frac{G_F^2}{8\pi} f_{D_s}^2 m_{\ell}^2 M_{D_s}
\left(1-\frac{m_{\ell}^2}{M_{D_s}^2}\right)^2 \left|V_{cs}\right|^2 r_{D_s}\;, \label{equ_rate}
\end{equation}
where in the MSSM one has
\cite{Hou:1992sy,Hewett:1995aw,Akeroyd:2007eh,Dobrescu:2008er,Akeroyd:2009tn}:
\begin{equation}
r_{D_s}\equiv \left[1+\left(\frac{1}{m_c+m_s}\right)\left(\frac{M_{D_s}}{m_{H^+}}\right)^2\left(m_c-
\frac{m_s\tan^2\beta}{1+\epsilon_0 \tan\beta}\right) \right]^2  \;.
\label{rs}
\end{equation}
Here $m_c$ and $m_s$ are the masses of the charm and strange quarks respectively, $M_{D_s}$ is the mass of the $D_s$ meson, $V_{cs}$ is a CKM matrix element, and $m_{\ell}$ is the lepton mass.   
We note that $\epsilon_0$ in Eq.~(\ref{rs}) is not the same as $\epsilon_0$ in Eq.~(\ref{eq:RMSSM}),
because they are functions of different SUSY parameters.
The term $m_s\tan^2\beta$, which originates from the strange quark Yukawa coupling, can give rise to a 
non-negligible suppression of $r_{D_s}$ for large values of $\tan\beta$. 
Note that the magnitude of the $H^\pm$ contribution depends on the ratio of quark masses $m_s/(m_c+m_s)$, and an analogous uncertainty is not present for the $H^\pm$ contribution to the decay $B_u\to  \ell\nu$.
However, there is now a very precise calculation (error $\sim 1\%$) of $m_s/m_c$ from unquenched LQCD \cite{Davies:2009ih}.

There are various unquenched lattice calculations of $f_{D_s}$ and the current situation is summarized in \cite{Rosner:2010ak}.
The value with the smallest quoted error is from the HPQCD Collaboration, $f_{D_s}=241\pm 3$ MeV \cite{Follana:2007uv} (with a provisional update of $f_{D_s}= 247 \pm 2 \mbox{ MeV}$ \cite{HPQCD_update}). 
The MILC collaboration obtains $f_{D_s}=249\pm 11$ MeV \cite{Aubin:2005ar}
(with a provisional update of $f_{D_s}=260\pm 10$ MeV \cite{Bazavov:2009ii}).
A partially quenched calculation by the ETMC Collaboration gives
$f_{D_s}=244\pm 8$ MeV \cite{Blossier:2009bx}.
 
On the experimental side, $D_s \to \tau\nu$ has been measured at CLEO-c 
for three decay modes of $\tau$ ($e\nu\nu, \pi\nu,\rho\nu$) 
\cite{Alexander:2009ux,Onyisi:2009th,Collaboration:2009tk}.
These are absolute branching ratio measurements. 
Moreover, a measurement of $D_s \to \tau\nu$ has been performed at
BABAR \cite{:2010qj} for one decay mode of $\tau$ ($e\nu\nu$).
Unlike the CLEO-c measurements above, this measurement is normalized to the  
branching ratio of the decay $D_s\to K^\pm K^0$.
The decay $D_s \to \mu\nu$ has been measured at CLEO-c \cite{Alexander:2009ux}, 
BELLE \cite{:2007ws} and BABAR \cite{Aubert:2006sd}. Both the 
CLEO-c and BELLE measurements are of the absolute branching ratio, while the
BABAR measurement is normalized to the decay $D_s\to \phi^0\pi^\pm$ 
(note that this differs from the decay $D_s\to K^\pm K^0$ which is used in the 
BABAR measurement of $D_s \to \tau\nu$).
Due to the sizeable uncertainty in the branching ratio of  $D_s\to \phi^0\pi^\pm$,
the BABAR measurement \cite{Aubert:2006sd} is not included in the 
averages given in \cite{Rosner:2010ak}.
The average of the six measurements 
\cite{Alexander:2009ux,Onyisi:2009th,Collaboration:2009tk,:2010qj,:2007ws} 
results in a world average of $f_{D_s}=257.5\pm 6.1$ MeV (derived in \cite{Rosner:2010ak}
and is dominated by the average of the four CLEO-c measurements).
The world average given in \cite{Barberio:2008fa} 
is lower ($f_{D_s}=254.6\pm 5.9$ MeV) because it includes the BABAR measurement of
$f_{D_s}$ in \cite{Aubert:2006sd} (although with a reinterpreted value, as explained in \cite{Schwartz:2009hv}).

The experimental results for the branching ratios are \cite{Barberio:2008fa,Akeroyd:2009tn}:
\begin{equation}
\mathrm{BR}(D_s \to \tau\nu_\tau)_\mathrm{exp}=(5.38\pm 0.32)\times 10^{-2}\;, \label{interval_dstaunu}
\end{equation}
\begin{equation}
\mathrm{BR}(D_s \to \mu\nu_\mu)_\mathrm{exp}=(5.81\pm 0.43)\times 10^{-3}\;, \label{interval_dsmunu}
\end{equation}
\begin{equation}
\mathrm{BR}(D \to \mu\nu_\mu)_\mathrm{exp}=(3.82\pm 0.33)\times 10^{-4}\;. \label{interval_dmunu}
\end{equation}
More precise measurements will be possible at BES-III.
A precision of a few percent ($2\%\to 4\%$) is expected for all three decays $D_s \to \tau\nu_\tau$,
$D_s \to \mu\nu_\mu$, and $D \to \mu\nu_\mu$, before a high-luminosity $B$ factory starts to
operate. In principle, data from the current $B$ factories (and at high-luminosity upgrades)
could also be used to provide further measurements of 
$D_s \to \mu\nu$ and $D_s \to \tau\nu$. The recent measurement of
 $D_s \to \tau\nu$ (with $\tau\to e\nu\nu$) by BABAR \cite{:2010qj} suggests that a similar 
measurement could be attempted at BELLE. Other decay modes of $\tau$ (e.g. $\pi^\pm\nu$ and $\rho^\pm\nu$, as done at CLEO-c) might also provide additional measurements.
We note that BES-III has the capability to measure the decay channel
$D_s\to K^\pm K^0$ (which is used for normalizing in \cite{:2010qj}) more precisely, which 
will reduce the error in the extracted branching ratio of $D_s \to \tau\nu_\tau$ 
from techniques like that used in \cite{:2010qj}.

The SM predictions for the BRs, obtained using SuperIso v2.8, are as follows:
\begin{eqnarray}
\rm{BR}(D_s \to \tau \nu_\tau)=(4.82 \pm 0.14) \times 10^{-2}\;, \\
\rm{BR}(D_s \to \mu \nu_\mu)=(4.98 \pm 0.14) \times 10^{-3}\;,\\
\rm{BR}(D \to \mu \nu_\mu)=(3.89 \pm 0.16) \times 10^{-4}\;.
\end{eqnarray}%
in which $f_{D_s}= 241 \pm 3 \mbox{ MeV}$ is used.

%%%%%%%%%%%%%%%%%%%%%%%%%%%%%%%%%%%%%%%%%%%%%%%%%%%%%%%%%%%%%%%%%%%%
%
\subsection{The decay $B_s\to \mu^+\mu^-$}
The decay $B_s \to \mu^+ \mu^-$ is  the only decay in the double ratio which has not been measured.
In the SM, the main theoretical uncertainty is from $f_{B_s}$ and $V_{ts}$, but the latter can be taken as a well-measured parameter by assuming CKM unitarity. 
It has been emphasized in many works \cite{Choudhury:1998ze,Babu:1999hn,Carena:2006ai,Ellis:2007ss,Mahmoudi:2007gd}
that the decay $B_s \to \mu^+ \mu^-$
is very sensitive to the presence of SUSY particles. 
At high $\tan\beta$, the MSSM contribution to this process is dominated by the exchange of neutral Higgs bosons. 
We therefore expect indirect constraints on $m_{H^+}$ and $\tan\beta$ from the MSSM mass relations. 
The ${\rm BR}(B_s\to \mu^+\mu^-)$ can be expressed as \cite{Bobeth:2001sq}
\begin{eqnarray}
\mathrm{BR}(B_s \to \mu^+ \mu^-) &=& \frac{G_F^2 \alpha^2}{64 \pi^3} f_{B_s}^2 \tau_{B_s} M_{B_s}^3 |V_{tb}V_{ts}^*|^2 \sqrt{1-\frac{4 m_\mu^2}{M_{B_s}^2}} \nonumber \\
&\times& \left\{\left(1-\frac{4 m_\mu^2}{M_{B_s}^2}\right) M_{B_s}^2 | C_S |^2 + \left |C_P M_{B_s} -2 \, C_A \frac{m_\mu}{M_{B_s}} \right |^2\right\} \;,
\end{eqnarray}
where the coefficients $C_S$, $C_P$, and $C_A$ parametrize different contributions. Within the SM, $C_S$ and $C_P$ are small, whereas the main contribution entering through $C_A$ is helicity suppressed. In the MSSM, both $C_S$ and $C_P$ can receive large contributions from scalar exchange. The $B_s$ decay constant $f_{B_s}=238.8\pm 9.5$ MeV \cite{Laiho:2009eu} 
constitutes the main source of uncertainty in this expression. 

 We define
\begin{equation}
r_{B_s} \equiv \frac{\mathrm{BR}(B_s \to \mu^+ \mu^-)_{\mathrm{MSSM}} }{\mathrm{BR}(B_s \to \mu^+ \mu^-)_{\mathrm{SM}}}\;.
\end{equation}
The SM prediction, obtained with SuperIso v2.8 is
\begin{equation}
\mathrm{BR}(B_s \to \mu^+ \mu^-)_\mathrm{SM}=(3.21\pm 0.29) \times 10^{-9}\;,
\end{equation}
while the current experimental limit, derived by the CDF collaboration, is \cite{Barberio:2008fa,CDF9892}:
\begin{equation}
\mathrm{BR}(B_s \to \mu^+ \mu^-) < 4.3 \times 10^{-8} \label{interval_bsmumu}
\end{equation}
at 95\% C.L. The experimental limit is thus still an order of magnitude away from the SM prediction, allowing for a substantial SUSY contribution. Including theoretical uncertainties,
 we compare the MSSM prediction to the upper limit at 95\% C.L.
\begin{equation}
\mathrm{BR}(B_s \to \mu^+ \mu^-) < 4.7 \times 10^{-8}\;.
\end{equation}

Three experiments at the LHC (LHCb, CMS and ATLAS) will have sensitivity to BR($B_s \to \mu^+ \mu^-$)
which is superior to that at the Tevatron.
At LHCb, a $3\sigma$ signal would be established for a SM-like rate for  BR($B_s \to \mu^+ \mu^-$)
after one year at design luminosity (2 fb$^{-1}$) \cite{Lenzi:2007nq}. 
Slightly inferior sensitivity is expected at CMS and ATLAS \cite{Smizanska:2008qm}, and
a $3\sigma$ signal would require three years at design luminosity (30 fb$^{-1}$).
The dominant systematic error will be from normalizing to specific decays of $B_u$ and $B$, with the two main sources of error arising from:
i) the uncertainty in the relative production rate of $B_s$ and $B_u/B$; ii) the error in the measurements of the branching ratios of $B_u/B$ which are used as normalization channels.
In \cite{LHCb_roadmap} this error is estimated to be $13\%$, thus making it difficult to establish (at a high confidence level) enhancements of up to a factor of 3 compared to the SM expectation.
An error of $\sim 13\%$ (not including statistical error) seems 
to be realistic for a measurement of
a SM-like  BR($B_s \to \mu^+ \mu^-$).
%%%%%%%%%%%%%%%%%%%%%%%%%%%%%%%%%%%%%%%%%%%%%%%%%%%%%%%%%%%%%%%%%%%%
%
\subsection{Experimental limit on the double ratio}

The double ratio in Eq.~(\ref{doub-rat}) can be rewritten as
\begin{equation}
 R = \frac{\eta}{\eta_{\mathrm{SM}}} \;, \label{ratio}
\end{equation}
where
\begin{equation}
\eta \equiv \left(\frac{\mathrm{BR}(B_s\to\mu^+\mu^-)}{\mathrm{BR}(B_u \to \tau \nu)}\right) \Big/ \left(\frac{\mathrm{BR}(D_s\to\tau\nu)}{\mathrm{BR}(D\to\mu\nu)}\right) \;. \label{ratio_eta}
\end{equation}
As explained in section 2.3, this ratio does not suffer from uncertainties from the decay constants, 
and the main theoretical uncertainty in the evaluation of $\eta_{\mathrm{SM}}$ $(= ( 2.27 \pm 0.54 )\times 10^{-7})$ comes from $V_{ub}$.
To determine the experimental limit on the ratio (\ref{ratio_eta}), we combine the limits on the individual branching fractions in Eqs.~(\ref{interval_btaunu}), (\ref{interval_dstaunu}), (\ref{interval_dmunu}) and (\ref{interval_bsmumu}). We use a Gaussian distribution for the constraints in Eq.~(\ref{interval_btaunu}), (\ref{interval_dstaunu}) and (\ref{interval_dmunu}), and a ``flat'' distribution for the upper limit in Eq.~(\ref{interval_bsmumu}). 
Combining statistically these experimental limits, we obtain, at 95\% C.L.
\begin{equation}
\eta < 2.09 \times 10^{-6} \;,
\end{equation}
which yields the upper limit for $R$ at 95\% C.L.
\begin{equation}
R < 10.0 \;, 
\label{R_doubrat}
\end{equation}
in which the uncertainty from $V_{ub}$ is taken into account. In our numerical analysis we use
Eq.~(\ref{R_doubrat}) to constrain the supersymmetric parameter space in two scenarios of the MSSM.

%%%%%%%%%%%%%%%%%%%%%%%%%%%%%%%%%%%%%%%%%%%%%%%%%%%%%%%%%%%%%%%%%%%%
%
\subsection{Attainable precision for $|V_{ub}|$ using the double ratio}
It is possible to estimate the future error in a measurement of $|V_{ub}|$ using the double ratio, assuming SM-like measurements for all four decays. The uncertainty in $|V_{ub}|$ will come from the error in the measurements of the four leptonic decays (i.e. the left-hand side 
of Eq.~(\ref{doub-rat})).

We first use the current experimental errors for each measured branching ratio, and consider a SM-like value for BR($B_s \to \mu^+ \mu^-$) with 30\% relative experimental error. Calculating $|V_{ub}|$ with a random Monte-Carlo generator, we find
\begin{equation}
|V_{ub}| = (3.58 \pm 0.71) \times 10^{-3} \;,
\label{vub-doubrat}
\end{equation}
which corresponds to about 20\% relative error. Although the error in Eq.~(\ref{vub-doubrat}) is larger than that
for the inclusive and exclusive measurements
in Eq.~(\ref{vub-inclusive}) and Eq.~(\ref{vub-exclusive}), the double ratio would provide an additional direct measurement of $|V_{ub}|$ with reasonably good precision.
Moreover, such a measurement of $|V_{ub}|$ would be complementary, because the 
theoretical error in Eq.~(\ref{vub-doubrat})
is much smaller than that in Eq.~(\ref{vub-inclusive}) and Eq.~(\ref{vub-exclusive}).

The future measurements of the branching ratios involved in the double ratio will have reduced errors, such as 2.5\% for $D$ and $D_s$ decays, 
and (in the optimistic case) 5\% for BR($B\to \tau\nu$), and 13\% for BR($B_s \to \mu^+ \mu^-$). 
With this set of errors, the uncertainty in the determination of $|V_{ub}|$ would  be reduced:
\begin{equation}
|V_{ub}| = (3.58 \pm 0.27) \times 10^{-3} \;,
\end{equation}
corresponding to 7.5\% relative error. Hence the double ratio has the potential to
provide a measurement of $|V_{ub}|$ which has a precision comparable (or even superior) to that for the current measurements of $|V_{ub}|$ in Eq.~(\ref{vub-inclusive}) and Eq.~(\ref{vub-exclusive}).
Moreover, the analogous double ratio with the decay $B_u\to \mu\nu$ would lead to an independent 
determination of $|V_{ub}|$, with an error which is comparable to that obtained from the double ratio with BR($B\to \tau\nu$). Combining the measurements of these two double ratios could further improve the precision of $|V_{ub}|$.

We note that the inclusive and exclusive determinations of $|V_{ub}|$ will also 
have reduced experimental errors at a high-luminosity $B$ factory \cite{Akeroyd:2004mj}, although 
the extent of the improvement of the theoretical error is not clear at present. The
inclusive determination of $|V_{ub}|$ used in \cite{:2009tp} is a very promising approach because 
it has a relatively small theoretical error, and so such a method is likely to lead to the most precise  measurement of $|V_{ub}|$ at a high-luminosity $B$ factory.
However, even if the total error from the conventional determinations of $|V_{ub}|$ is smaller than that 
from the double ratios, the complementary measurements of $|V_{ub}|$ from the latter will be important additional information, especially if the inclusive and exclusive determinations continue to give significantly different central values (which is the case at present).

%%%%%%%%%%%%%%%%%%%%%%%%%%%%%%%%%%%%%%%%%%%%%%%%%%%%%%%%%%%%%%%%%%%%
%
\section{Constraints on the supersymmetric parameter space}

In this section we study the dependence of the double ratio $R$ on the parameter space of supersymmetric models.
We consider two scenarios in the MSSM: i) the constrained MSSM (CMSSM) and ii) the Non-Universal Higgs Mass model (NUHM), with minimal flavour violation (MFV). 
The CMSSM is characterized by a set of universal parameters at 
the GUT scale $\lbrace m_0$, $m_{1/2}$, $A_0$, $\tan\beta\rbrace$, as well as the sign of the $\mu$ parameter, while in the NUHM the universality of the GUT scale mass parameters is relaxed for the Higgs sector leading to two additional parameters, $\mu$ and $m_A$. This additional freedom implies in particular that the charged Higgs mass can be considered as a free parameter.

To investigate the parameter spaces of the CMSSM and NUHM we generate 300,000 random points scanning over the ranges $m_0 \in [50,2000]$ GeV, $m_{1/2} \in [50,2000]$ GeV, $A_0 \in [-2000,2000]$ GeV, $\tan\beta \in [1,60]$ with positive $\mu$ in CMSSM and $\mu \in [-1000,2000]$ GeV and $m_A \in [5,2000]$ GeV in NUHM. For each point we calculate the spectrum of SUSY particle masses and couplings using SOFTSUSY 3.1 \cite{Allanach:2001kg} and we compute the branching fractions $\rm{BR}(B_u\to \tau\nu_\tau)$, $\rm{BR}(D \to \mu \nu_\mu)$, $\rm{BR}(D_s \to \tau \nu_\tau)$, $\rm{BR}(D_s \to \mu \nu_\mu)$ and $\rm{BR}(B_s\to \mu^+\mu^-)$ using SuperIso v2.8 \cite{Mahmoudi:2007vz,Mahmoudi:2008tp}. The obtained values are then used to calculate the double ratio.

\begin{figure}[!t]
\begin{center}
\includegraphics[width=8.cm]{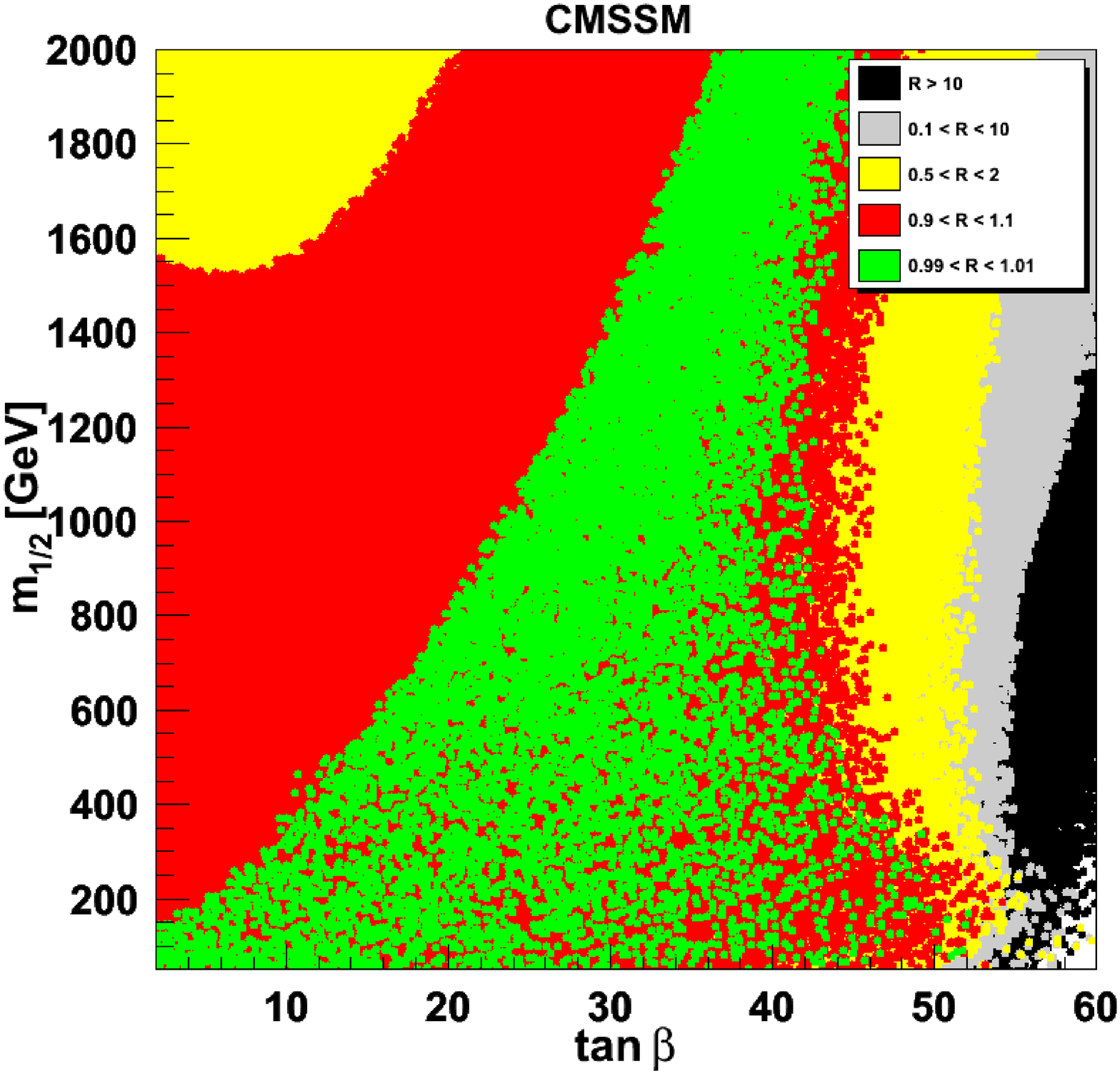}\includegraphics[width=8.cm]{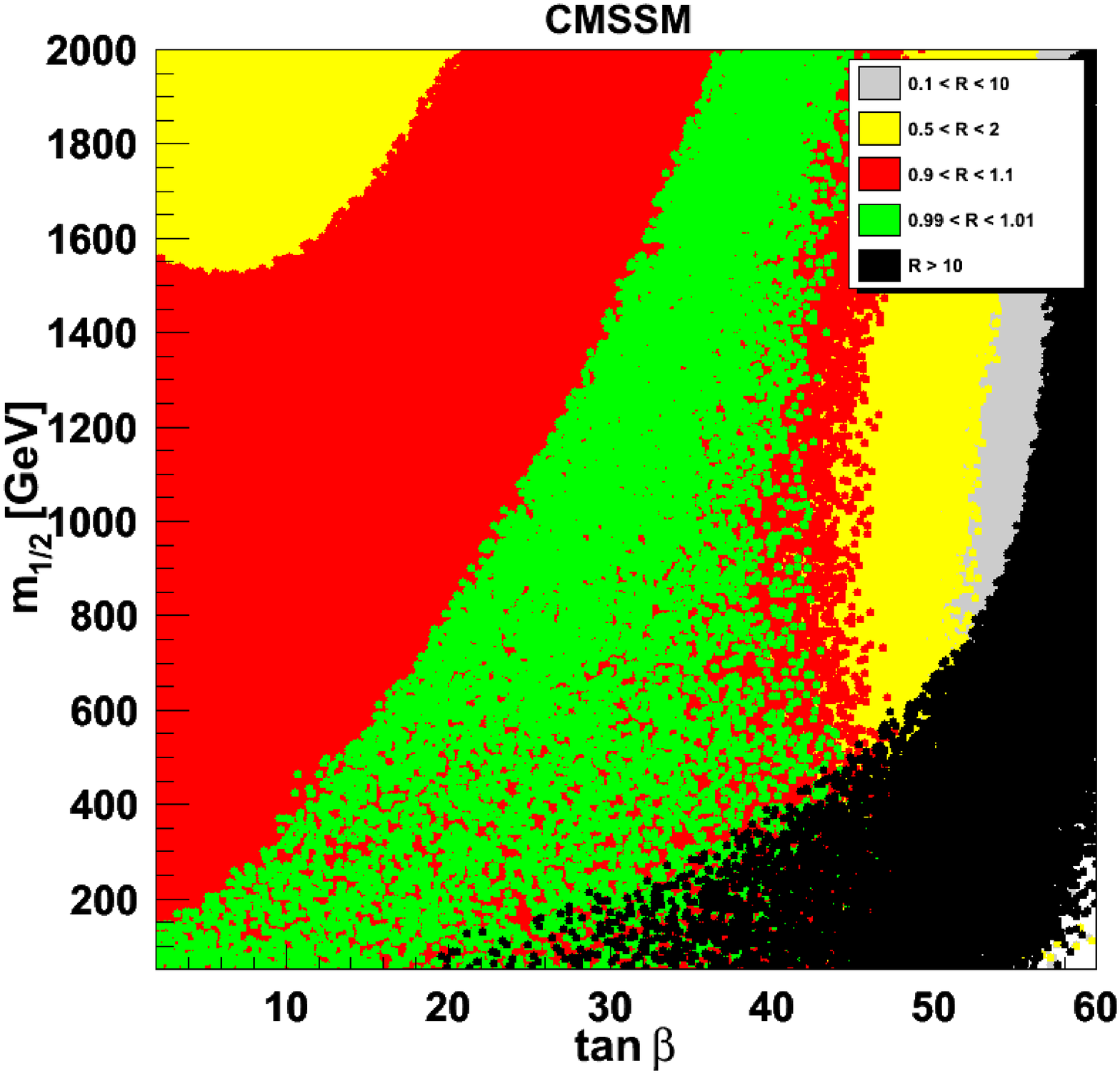}
\caption{Variation of the double ratio $R$ in the CMSSM plane $(\tan\beta,m_{1/2})$. The zones in green and red delimit 1\% and 10\% deviation from the SM value respectively. In the yellow zone, $R$ can be a factor of 2 away from the SM and in the grey zone
by a factor of 10. The black points are excluded at $95\%$ C.L.}
\label{fig1_msugra}
\end{center}
\end{figure}

In Fig.~\ref{fig1_msugra}, the results are shown for CMSSM in two separate plots. The scan over the four dimensional parameter space is projected into the plane $(\tan\beta,m_{1/2})$, where different colours correspond to different intervals for $R$. The points in black are excluded by $R$. In the left plot, the excluded points are displayed in the background, while they are shown in the foreground in the right plot. The excluded region in the left plot is therefore independent of the other SUSY parameters. In a large part of the parameter space, the double ratio is SM-like. In these regions, $|V_{ub}|$ can be determined with almost no additional deviation due to SUSY. On the other hand, the area with $\tan\beta > 55$ and $m_{1/2} < 1000$ GeV is excluded with no dependence on the lattice inputs. 

\begin{figure}[!t]
\begin{center}
\includegraphics[width=8.cm]{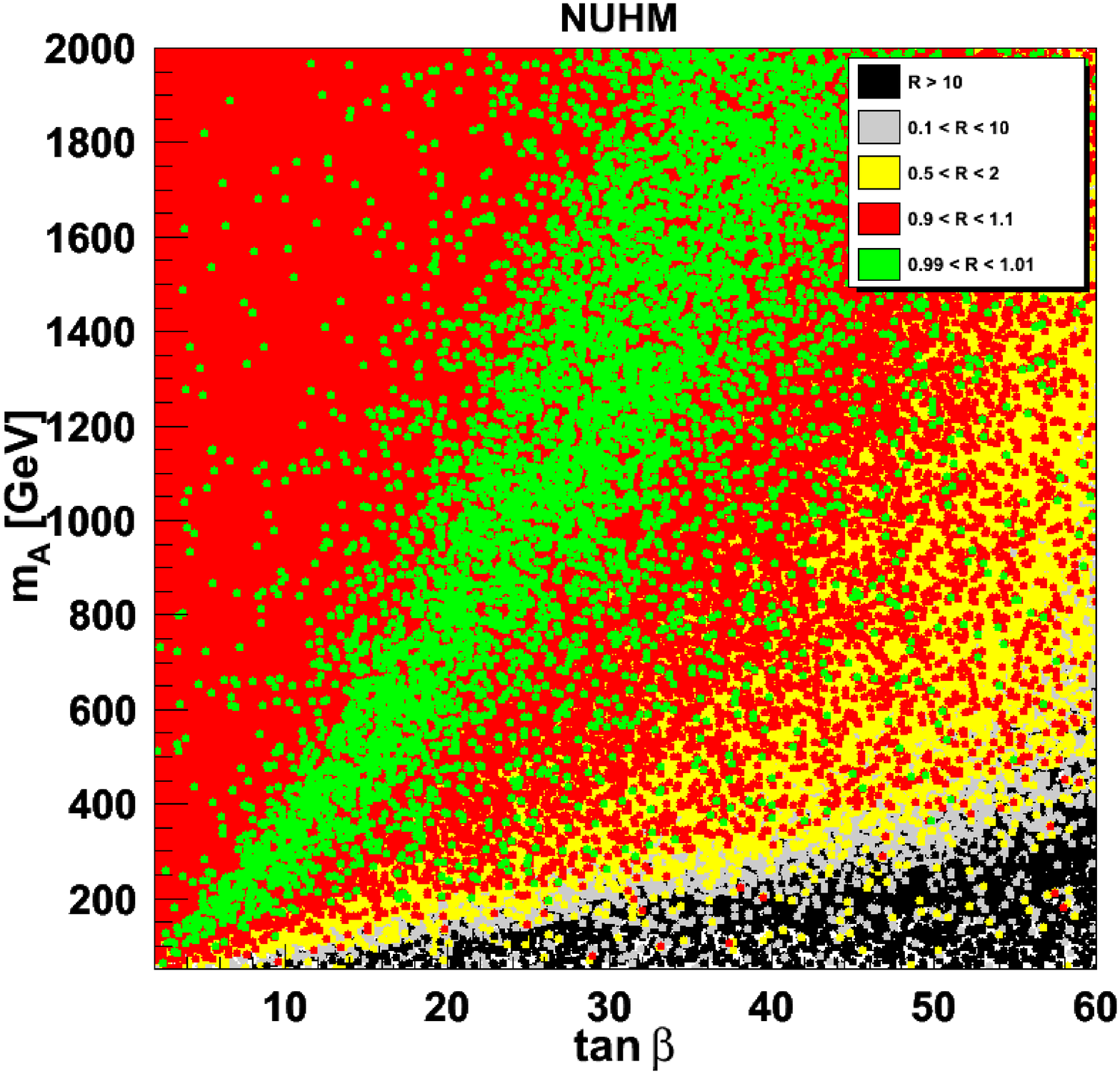}\includegraphics[width=8.cm]{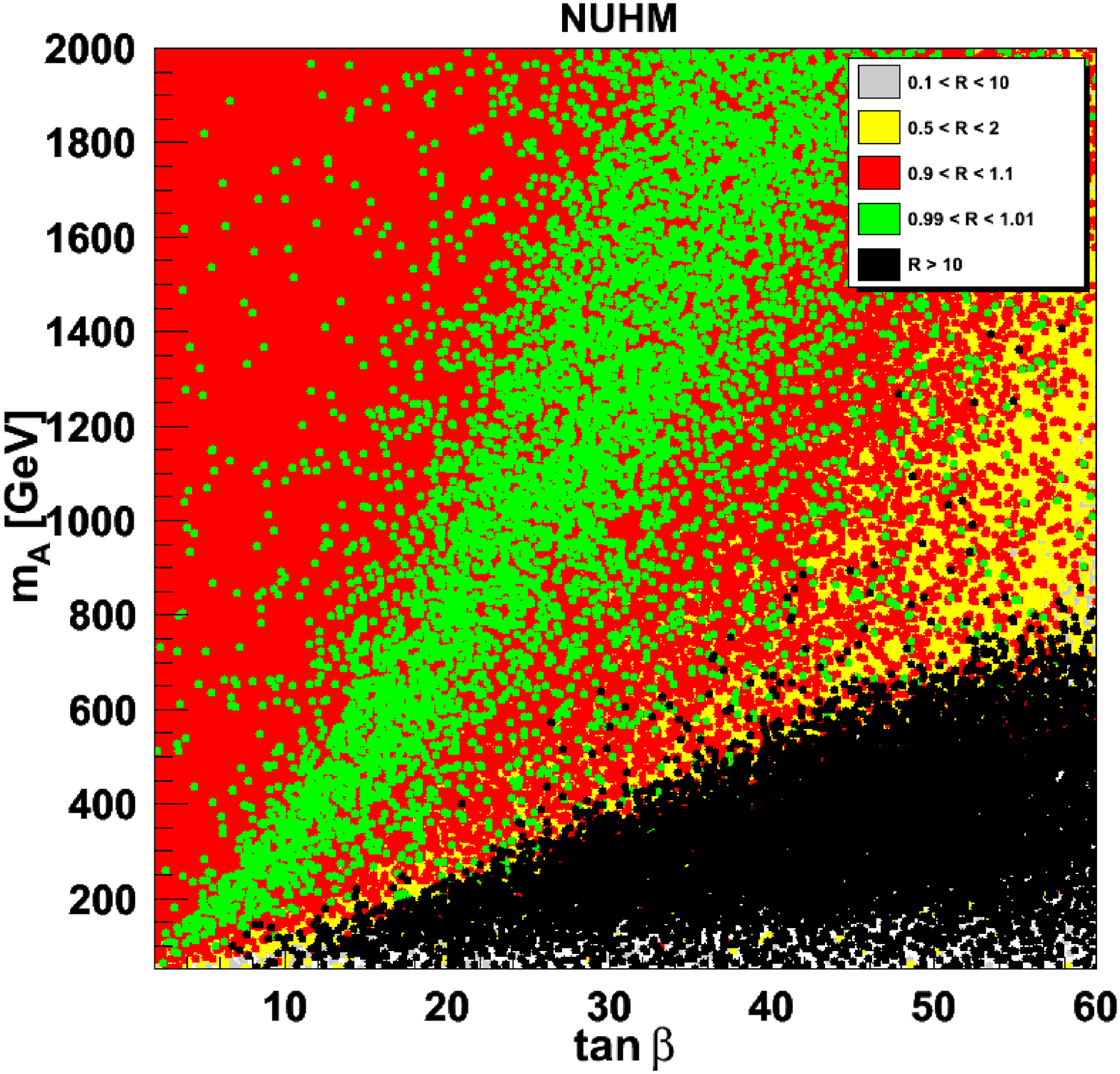}
\caption{Variation of the double ratio $R$ in the NUHM plane $(\tan\beta,m_A)$. The zones in green and red delimit 1\% and 10\% deviation from the SM value respectively. In the yellow zone $R$ can be a factor of 2 away from the SM and in grey zone by a factor 10. The black points are excluded at $95\%$ C.L.}
\label{fig2_nuhm}
\end{center}
\end{figure}

In a similar way, in Fig.~\ref{fig2_nuhm} the results are shown for NUHM in the plane $(\tan\beta,m_A)$. Again, the points in black are excluded by $R$, in the background in the left plot and in the foreground in the right plot. As in CMSSM, in a large part of the parameter space, the double ratio is SM-like, while the region $m_A/\tan\beta \lesssim 8$ GeV is excluded with no dependence on the lattice inputs. 


\begin{figure}[!t]
\begin{center}
\includegraphics[width=8.cm]{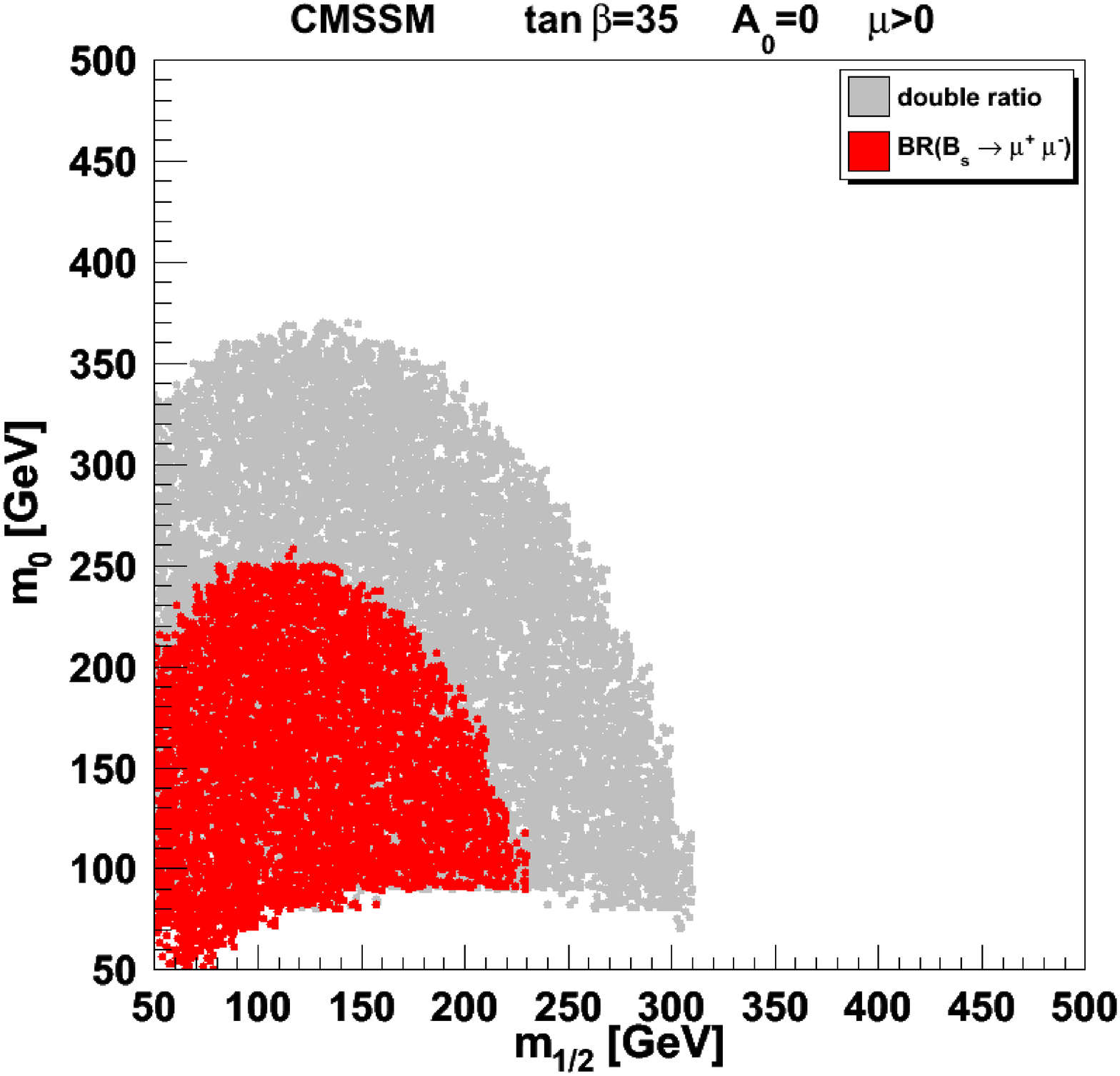}\includegraphics[width=8.cm]{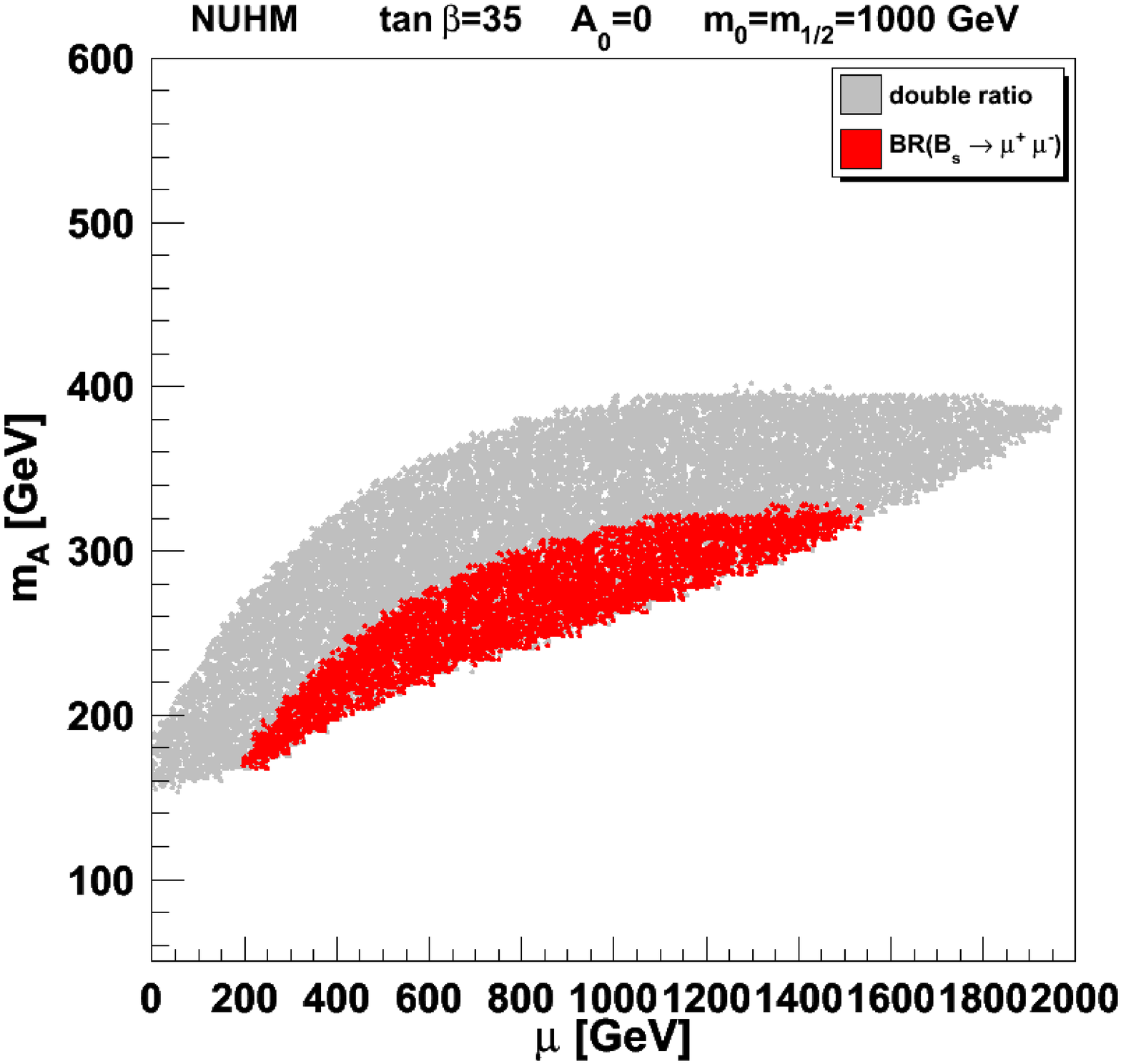}
\caption{In the left panel, CMSSM parameter plane $(m_{1/2},m_0)$ with $\tan\beta = 35$, $A_0=0$ and $\mu >0$.
In the right panel, NUHM parameter plane $(\mu,m_A)$ with $\tan\beta = 35$, $A_0=0$ and $m_0=m_{1/2}=1000$ GeV. 
The points in red are excluded at $95\%$ C.L. by $\mathrm{BR}(B_s\to\mu^+\mu^-)$ and in grey by $R$.}
\label{fig3}
\end{center}
\end{figure}

In order to compare the constraints on the SUSY parameter space from the double ratio with those 
from $\mathrm{BR}(B_s\to\mu^+\mu^-)$ alone, in Fig.~\ref{fig3} we show examples of 
parameter space in the CMSSM and in the NUHM which are excluded from both observables. 
It can be seen that the double ratio, being a combination of four different constraints, can be
more constraining than the branching ratio of $B_s\to\mu^+\mu^-$ taken individually. 
Importantly, contrary to $B_s\to\mu^+\mu^-$, the double ratio does not depend on lattice inputs, but instead depends on $|V_{ub}|$, whose magnitude is already constrained from various distinct experimental methods (i.e. direct measurements and global-SM fits).
Hence the double ratio is an important alternative to $\mathrm{BR}(B_s\to\mu^+\mu^-)$ as a probe of 
the SUSY parameter space.
 
\begin{figure}[!t]
\begin{center}
\includegraphics[width=8.cm]{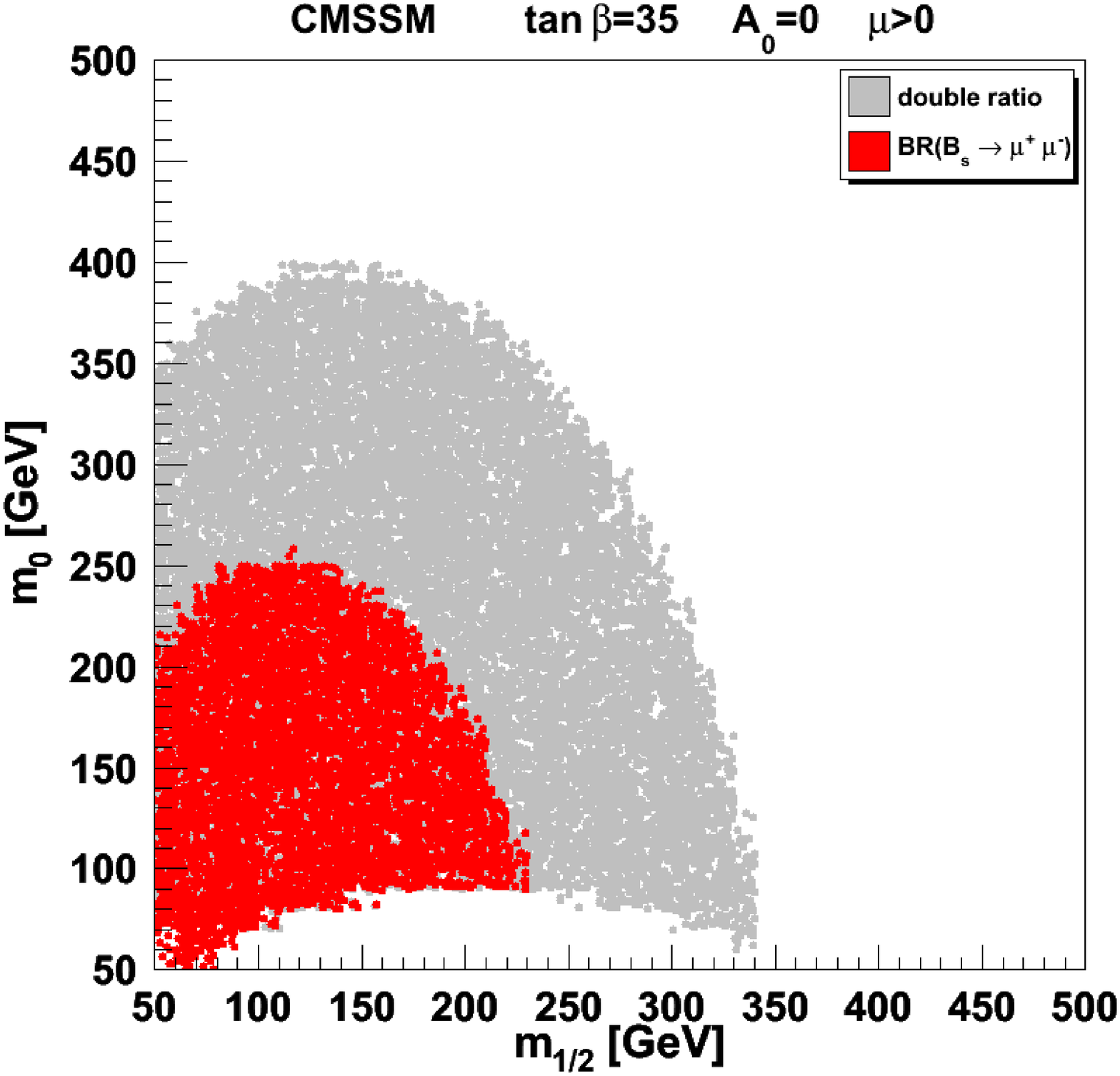}\includegraphics[width=8.cm]{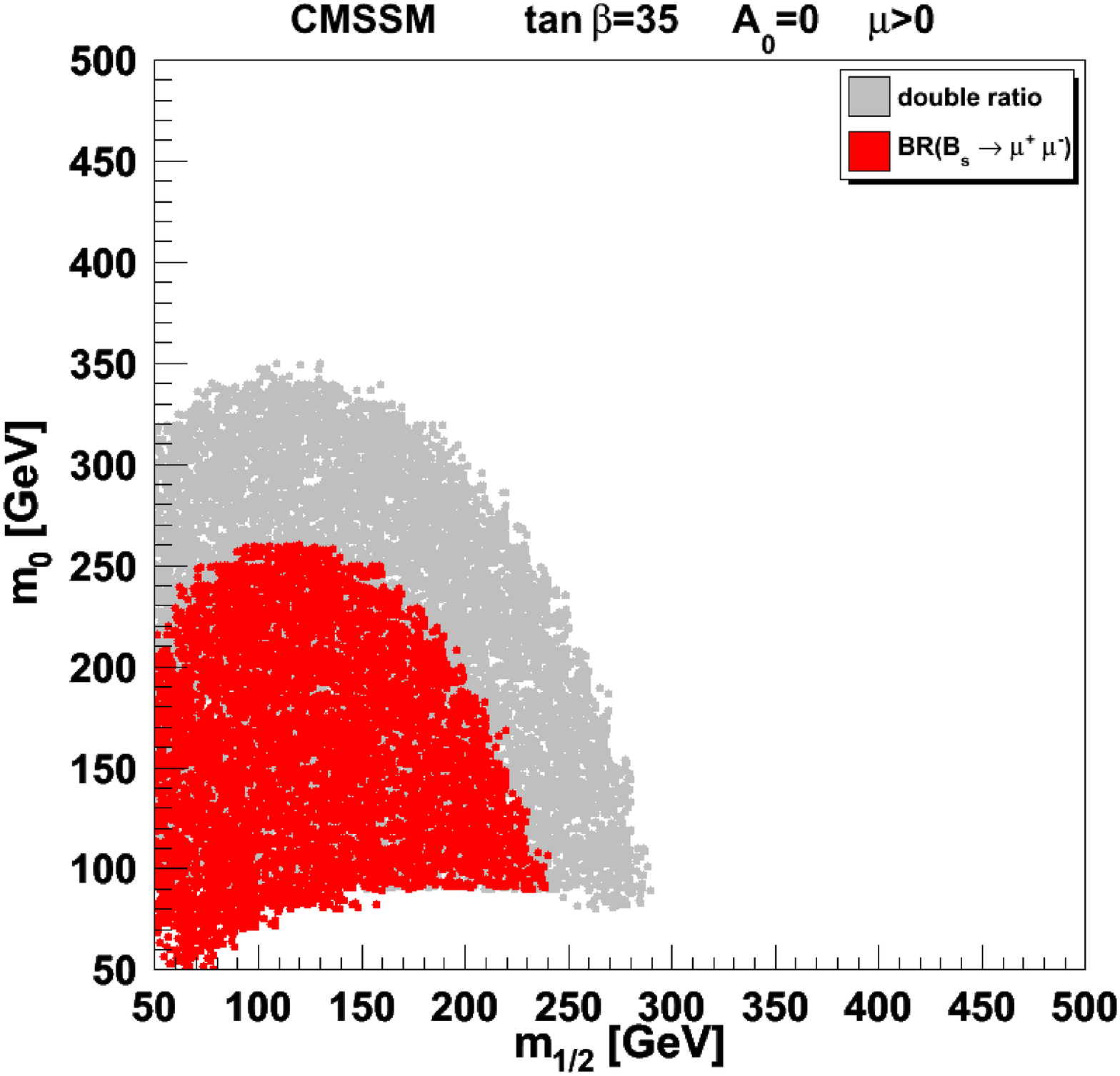}
\caption{In the left panel, CMSSM parameter plane $(m_{1/2},m_0)$ with $\tan\beta = 35$, $A_0=0$ and $\mu >0$ for $|V_{ub}|=3.48 \times 10^{-3}$
and $f_{B_s}=238.8$ MeV. In the right panel, $|V_{ub}|=4.5 \times 10^{-3}$ and $f_{B_s}=250$ MeV. The points in red are excluded at $95\%$ C.L. by BR($B_s\to\mu^+\mu^-$) and in grey by $R$.}
\label{fig4}
\end{center}
\end{figure}

The conclusion that the double ratio can be more powerful at constraining SUSY parameters than BR($B_s\to\mu^+\mu^-$) is irrespective of the freedom in choosing the input parameters 
(which are $|V_{ub}|$ for the double ratio and $f_{B_s}$ for BR($B_s\to\mu^+\mu^-$)).
This is shown in Fig.~\ref{fig4} where both the least favourable (with high $|V_{ub}|$ and high $f_{B_s}$) and most favourable (with low $|V_{ub}|$ and low $f_{B_s}$) cases for the double ratio 
are considered. In both cases, the double ratio excludes a greater region of the SUSY parameter space.


%%%%%%%%%%%%%%%%%%%%%%%%%%%%%%%%%%%%%%%%%%%%%%%%%%%%%%%%%%%%%%%%%%%%
%
\section{Conclusions}

Ongoing and forthcoming experiments will significantly improve the precision of the measurements of
the leptonic decays $B_u\to \tau\nu$ (at a high-luminosity $B$ factory), $D\to \mu\nu$ 
(at BES-III) and $D_s\to \mu\nu/\tau\nu$ (at a high-luminosity $B$ factory and BES-III).
A first measurement of the decay $B_u\to \mu\nu$ will be possible in the early stages of operation of
a high-luminosity $B$ factory if its branching ratio is comparable to the prediction in the Standard Model.
The ongoing experiments LHCb, CMS and ATLAS will provide the first measurements of the decay $B_s\to \mu^+\mu^-$ if its branching ratio is comparable to (or greater than) the prediction in the Standard Model. The branching ratio of each of these leptonic decays depends on the magnitude of the relevant decay constant, which can be calculated with techniques such as lattice QCD.  
Double ratios involving four of the decays can be defined in which this dependence on the decay constants is essentially eliminated, thus giving rise to an alternative observable with substantially reduced theoretical error.

A measurement of the double ratio involving $B_u\to \tau\nu$, $D\to \mu\nu$, $D_s\to \mu\nu/\tau\nu$ and $B_s\to \mu^+\mu^-$
would enable a complementary measurement of the CKM matrix element $V_{ub}$ \cite{Grinstein:1993ys,Ligeti:2003hp} with a theoretical error which is much smaller than that
present in the conventional approaches to measure $|V_{ub}|$ (which utilize inclusive and exclusive semi-leptonic decays of $B$ mesons).
We quantified the experimental error in a possible future measurement of $|V_{ub}|$ using the above double ratio, and showed that it can be competitive with the anticipated precision from the conventional approaches. Such an additional measurement of $|V_{ub}|$ would be particularly beneficial if the current disagreement between the inclusive and exclusive determinations of $|V_{ub}|$ persists
into the era of a high-luminosity $B$ factory.

In the context of supersymmetric models the above leptonic decays are also mediated by New Physics particles.
We showed that such double ratios can be more effective than the individual leptonic decays as a probe of the parameter space of supersymmetric models. We emphasized that the double ratios have the advantage of using $|V_{ub}|$ as an input parameter (for which there is 
experimental information), while the individual decays have an uncertainty from the decay constants (e.g. $f_{B_s}$), and hence a reliance on theoretical techniques such as lattice QCD.
Consequently, the double ratios of leptonic decays are an alternative and competitive probe of the parameter space of supersymmetric models.

%%%%%%%%%%%%%%%%%%%%%%%%%%%%%%%%%%%%%%%%%%%%%%%%%%%%%%%%%%%%%%%%%%%%
%
\section*{Acknowledgements}
FM is grateful to St\'ephane Monteil for useful discussions. AGA was supported by the ``National Central University Plan to Develop First-class Universities''.  

%%%%%%%%%%%%%%%%%%%%%%%%%%%%%%%%%%%%%%%%%%%%%%%%%%%%%%%%%%%%%%%%%%%%
%

\end{document}